# Production and measuring methods and procedures in precision optical cavities production

Jiří BENEŠ[1], František PROCHÁSKA[1], Zdeněk RAIL[1], David TOMKA[1], Lenka PRAVDOVÁ[2], Ondřej ČÍP[2]

[1] Regional Center for Special Optics and Optoelectronic Systems (TOPTEC), Institute of Plasma Physics, Academy of Sciences of the Czech Republic, Za Slovankou 1782/3, 182 00 Prague 8, Czech Republic

[2]Institute of ScientificInstruments, Academy of Sciences of the Czech Republic, Královopolská 147, 612 64Brno, Czech Republic

## Abstract

This work presents the development of the production process of a Fabry–Perot type laser resonator. It describes the acquired knowledge in the field of production of very precise optical standards, cavities, and lenses. In addition, it describes the measurement methods used in production. The work is intended for industry and science.

## Introduction

A Fabry–Perot type optical resonator is a very effectivefilter of optical frequencies. It consists of two mirrors and a spacer determining their distance, which defines the fundamental resonant frequency. By combining anoptical radiation coherent source and a high–quality resonator, a highly coherent laser source can be obtained using a feedback loop. The optical frequency stability of the laser source is directly proportional to the thermomechanical stability of the optical resonator spacer material. A special category is represented by high–quality optical resonators designed for the assembly of so–called super coherent lasers, which are characterized by a very narrow free spectral range in the order of units or fractions of Hz [1]. High–quality resonators are used to maintain a sufficiently stable optical frequency of the laser in cooled–atom or atomic–ion spectroscopy, e.g., in the optical atomic clock implementation [2]. Super coherent lasers are an essential part of phase–coherent distribution optical fiber links, which are used to transmit accurate frequency and time over long distances [3].

## Solutionsuggestion

Corning ULE material was selected for production during the design process. It is a titanium–silicate glass custom–made forthe application. The material is characterized by very low thermal expansion around the CTE point [4]. The first important parameter of the system is the working temperature. Maintaining a constant ambient temperature by cooling is a very difficult task. The whole application was thus shifted to a temperature of 31 °C. This temperature was maintained by a very precise heating system. The ULE material in this application significantly reduces the temperature drift according to the formula

$$\frac{\Delta v}{v} = \frac{\Delta l}{l} \sim 10^{-9}(T - T_c)^2.$$

A thermally stabilized resonator could have an average drift of 3.2 mHz/s and daily frequency deviations of up to 1.5 kHz [5].

Both mirrors in the assembly are madefrom a fused silica material. This material was chosen because of the combination of suitable optical properties, good machinability, and relatively low thermal expansion. Fused silica has a much higher thermal expansion than ULE. The expanding mirror thus deforms the spacer. Assuming that the optical connection of the mirror and the spacer is perfectly strong, this would affect the shape of the optical surface and the function of the whole resonator. As a result, in an extreme case,the aforementioned deformationsmay separatethe mirror from the spacer. The solution to this problem is to add a ULE ring to the exterior of the mirror [6]. Thanks to this modification, a large part of the mirror becomes mechanically stabilized, which also results in a significant stabilization of the optical surfaces.

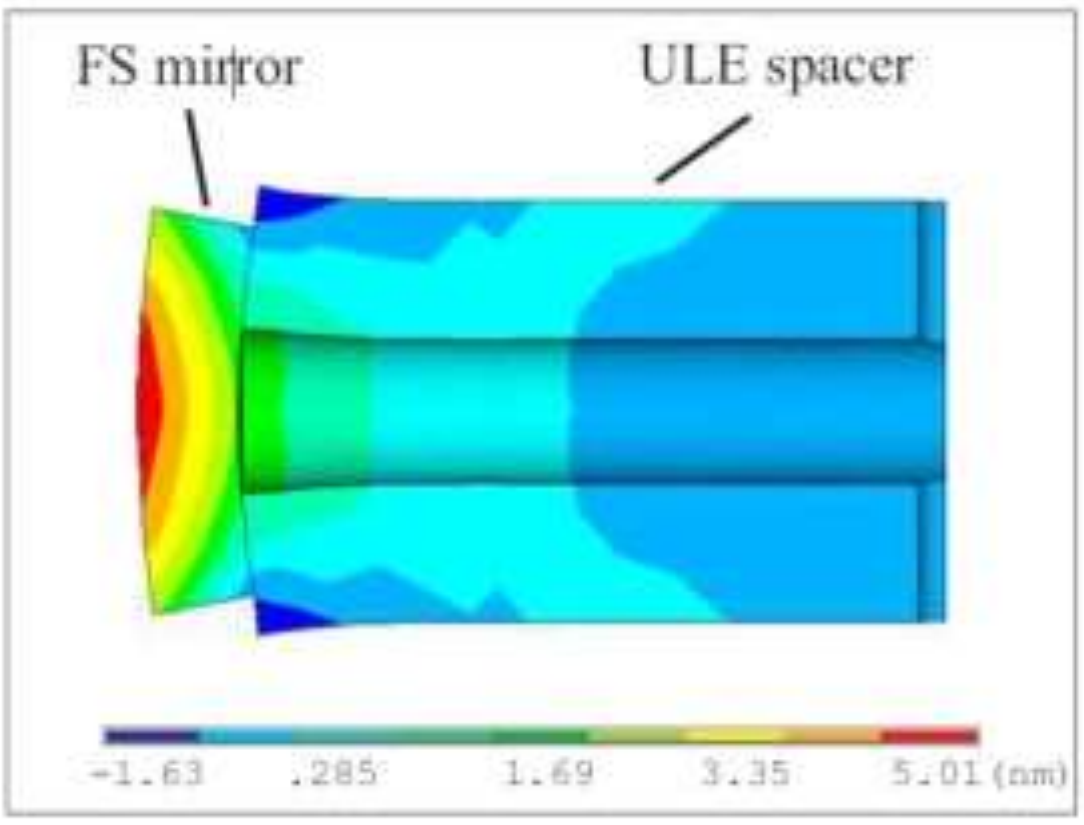

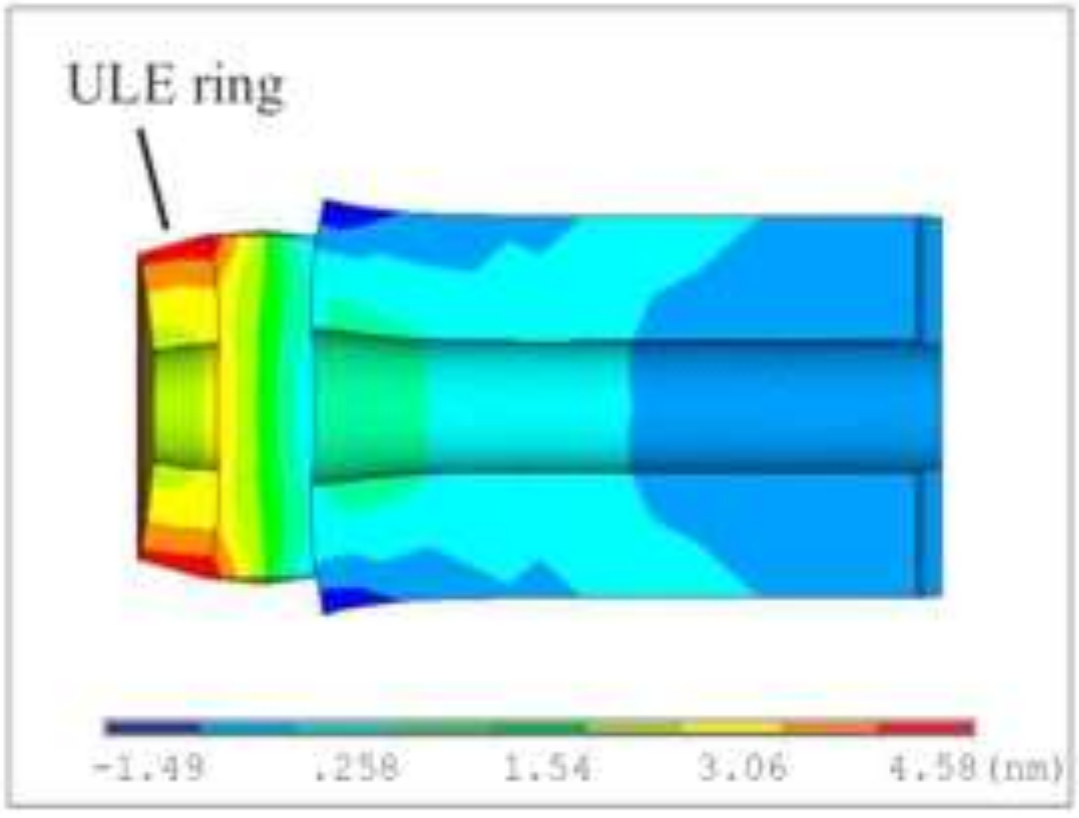

Figure 1. FEM simulations of the elastic cavity deformation after a 1 K temperature step: (a) FS mirror optically contacted to a ULE spacer, (b) additional ULE ring on the back side of the FS mirror to suppress its axial bending. Due to cylindrical symmetry, only a quarter of the cavity is simulated. Color scales how the axial displacement is. Source:[6]

The next problem addressed is the storage of the spacer. A part with a milled pocket was designed as a stand. Other published solutions were mounted on pedestals using Viton balls.

This solution requires the pocket to be milled into the spacer. With regard to the stresses that mechanical interventions create in the glass materials, another solution was chosen. The spaceris placed in a stand through Viton O-rings. The distance between the rings was determined to be 60 mm by simulating the bending of the resonator body.

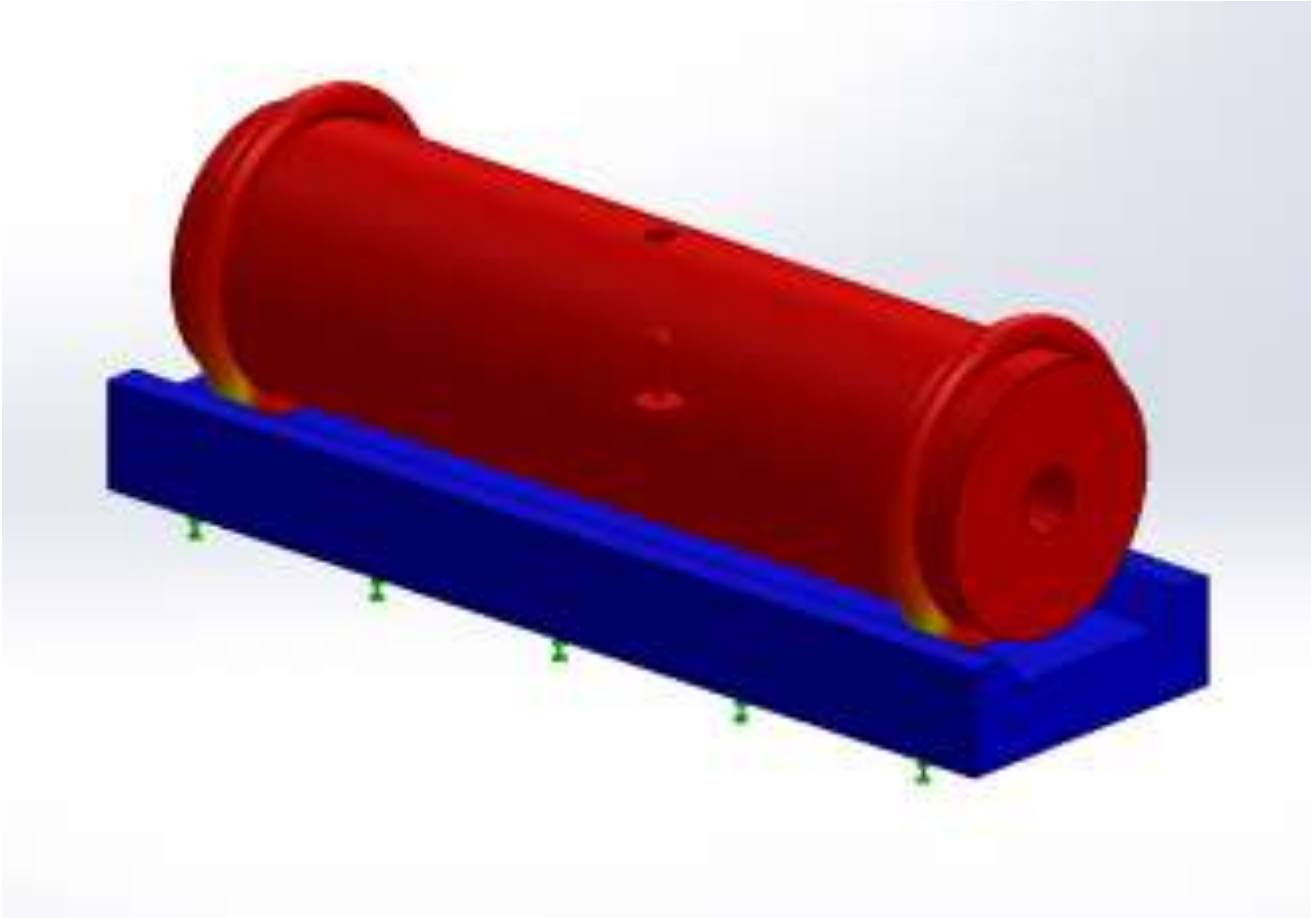

Figure2. Simulation of spacer placement on Viton O-rings

## Manufacturing processes

A prism with dimensions 181x53x53 mm was made. The faces of this prism were further ground so that their angular deviation did not exceed 5 arcsec. Such a level of precision was needed to achieve high accuracy in the subsequent drilling of the cavity. Because the cavity had an internal diameter of only 12 mm and a length of 181 mm, it was necessary to develop a working procedure. For this task, a 300 mm long hollow drill with a diamond bond was made to order. Several different problems were dealt with during the development. The first issue was the breaking off of the residual material inside the drill. In common applications of hollow drills, this problem does not occur because there is no arbitrary breaking off at a low depth of drilling. Normally, this problem only occurs when the length exceeds the diameterseveral times; in our application, the length exceeded the diameter15 times. Mechanical breakage by means of a lever proved to be the only solution. Another problem was keeping the borehole straight. It was necessary to perform experiments to determine the correct drilling speed to preventthe drill from bending. Finally, the method of monitoring and measuring with a laser ruler was used. The speed itself also depended, among other things, on the length of the residual glass in the drill and the depth at which the drill was currentlylocated. The issue of cooling was solved byinternal cooling of the drill. The cavity had to be measured and the cylindricality evaluated. After a thorough inspection, very precise centering holders were glued into the cavity and the prism was rounded to a diameter of 50 mm. This achieved sufficient coaxiality between the cavity and the outer cylinder.

All the previous steps can be classified as roughing steps. These processes can create local stresses in the material, which is undesirable in such a precise application. The spacer was therefore immersed in a 1: 1 mixture of hydrochloric acid and hydrofluoric acid. This so–called etching step removed the subsurface tension and also removed any waste residues from previous production processes.

In the next step, one of the fronts was furnished with attachments while the other one was glued to the cavity. Another ring–shaped attachment was glued to the outer cylinder. The total diameter of the area was then 140 mm. The surface was subsequently ground and angled to the axis of the cavity. Iterations of grinding and measurement on a 3D measuring machine (CMM) were used. The result of this procedure was an angular deviation of 3 arcsec. In the next step, the surface was polished with a Cerox suspension and a resin polishing tool. At the front of the spacer, an accuracy of λ/20 RMS and a micro-roughness of 0.55 nm were achieved. These values were sufficient for the subsequent vacuum connection.

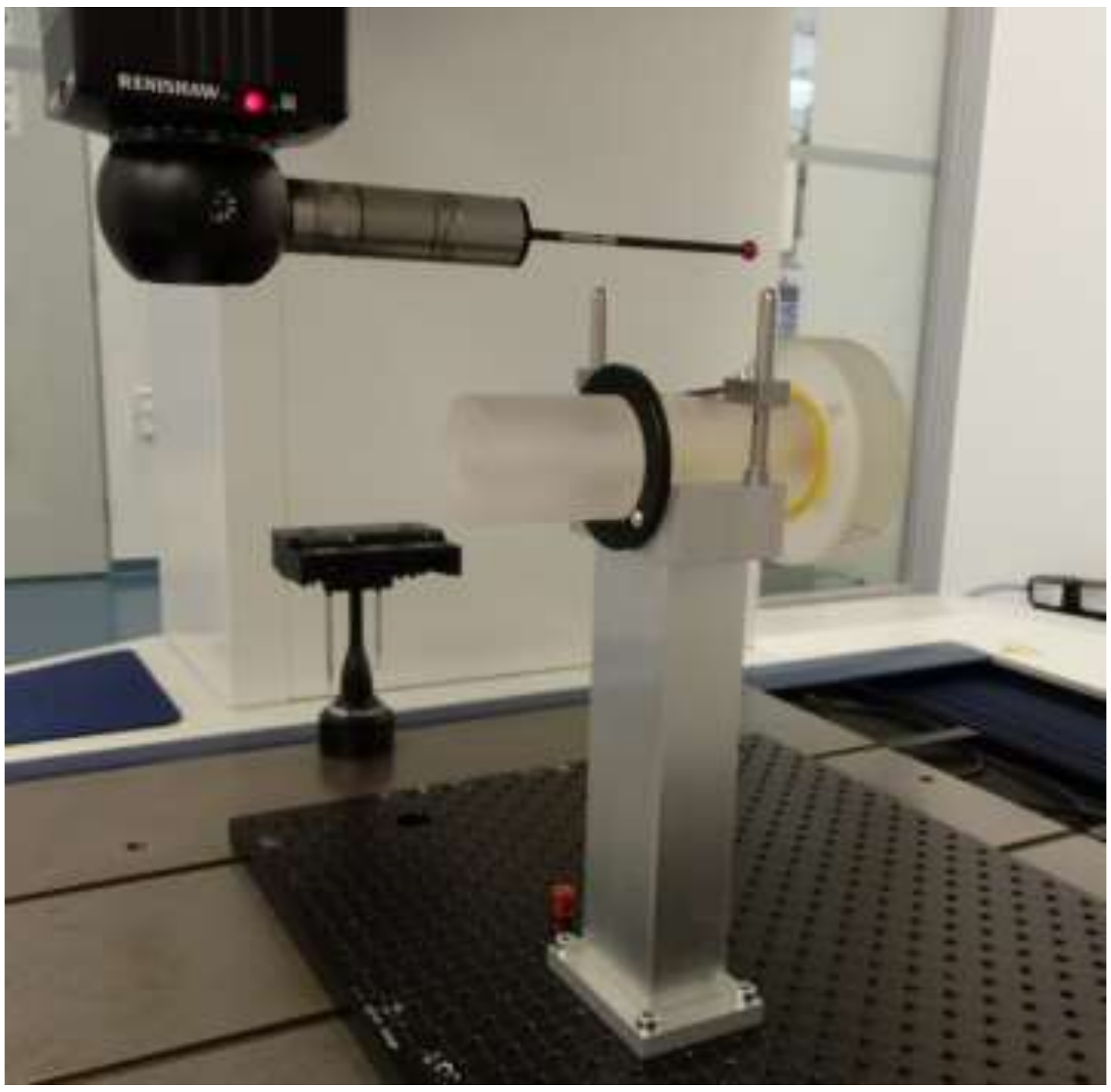

Figure 3. Clamping system in a Mitutoyo Legex 744 measuring machine

After the completion of the first surface, the attachments were glued to the other face. The area was ground and measured on a CMM. A wedging of the end face towards the other face was observed. After several iterations of measurements and grinding, a deviation of 5 arcsec was achieved. Subsequently, the surface was polished. During the polishing, the wedge was maintained to the required specifications by goniometer measurements.

Mirrors are another very important part of the resonator. The resonator is equipped with one planar and one concave mirror with a radius of 900 mm. The mirrors are fitted with compensating rings on the outside. The mirrors are made of fused silica. First, a flat surface on eachside is ground and polished on the mirrors. Grinding and polishing of flat surfaces takes place in lots of twenty. The mirrors are glued to the board and ground. First, a plane is generated on the CNC machine. Subsequently, the mirror plate is ground with a free SiC abrasive with a roughness of 600, 800, and 1000, with removals of 0.2, 0.2 and 0.1 mm.

In the next step, the board is polished with a resin tool and a polishing Cerox suspension. The mirrors are polished to a shape specification of 20 nm RMS. This procedure is applied to both sides of each mirror. During the production of opposite side, the plane parallelism of both sides is further monitored. After both planar surfaces are polished on all mirrors, several pieces are selected to produce a concave surface.

Several production methods were tested during manufacturing. The production of a concave surface with a curvature of 900 mm and a diameter of 12 mm is quite problematic. It is so shallow that commonly used procedures cannot be used. The manufacturing process was, therefore, modified so that an attachment was glued to the plane mirror.A concave surface was ground into this attachment. It was then ground with a free SiC abrasive and polished with a Cerox polishing suspension. After removing the attachment, a concave surface with a diameter of 12 mm remained in the middle of the planar surface.

During production,a problem arose with the fastening of the attachment. An optical connection was tested first, i.e., the application of two very precise planar surfaces which adhere to each other. The established connection is very strong but also very prone to liquid penetration and thus to collapsing, which is what happened in our experiment. Therefore, bonding with pitch was tested. Two binders were tested; onewith a melting point of 50 °C and another one with a melting point of 80 °C. During polishing, it was observed that the first binder could not withstand the heat generated during polishing. Finally, a second binder with a melting point of 80 °C was used. Tests revealed that both the concave and the planar surfaces were damaged at their boundary, which is critical, especially for the concave surface. To prevent the damage from interfering with the optical diameter of the concave surface, the design was modified and the concave surface was produced with a diameter of 16 mm. Achieving the lowest possible micro-roughness is very important for both mirror surfaces. With this parameter in mind, both surfaces were completed manually. The resulting parameters of the manufactured parts are recorded in Table 1.

Table1:Parameters of manufactured parts

| Part | Dimension | Nominal | Tolerance value | Production piece |
|---|---|---|---|---|
| Spacer | **Length** [mm] | 185 | 0.1 | 185.068 |
| | **Outer diameter** [mm] | 50 | 0.1 | 50.020 |
| | **Inner Diameter** [mm] | 11.6 | 0.1 | 11.551 |
| | **Inner Hole Cylindricality** [mm] | | 0.05 | 0.014 |
| | **Concentricity of both ends of the inner hole** [mm] | | 0.1 | 0.02 |
| | **Right front flatness**[nm RMS] | | 20 | 19 |
| | **Left front flatness** [nm RMS] | | 20 | 18 |
| | **Planparallelism of faces**[arcsec] | | 5 | 5 |
| | **Right front micro-roughness**[nm Sq] | | 0.6 | 0.498 |
| | **Left front micro-roughness** [nm Sq] | | 0.6 | 0.556 |
| | | | | |
| **Part** | **Dimension** | **Nominal** | **Tolerance value** | **Production piece** |
| Flat mirror | **Diameter** [mm] | 32 | 0.1 | 31.980 |
| | **Left surface flatness** [nm RMS] | | 20 | 17 |
| | **Right surface flatness** [nm RMS] | | 20 | 16 |
| | **Surface parallelism** [arcsec] | | 5 | 4 |
| | **Left surface micro-roughness** [nm Sq] | | 0.6 | 0.512 |
| | **Right surface micro-roughness** [nm Sq] | | 0.6 | 0.501 |
| Concave mirror | **Diameter** [mm] | 32 | 0.1 | 32.098 |
| | **Left surface flatness**[nm RMS] | | 20 | 18 |
| | **Right surface flatness** [nm RMS] | | 20 | 20 |
| | **Surface parallelism** [arcsec] | | 5 | 3 |
| | **Left surface micro-roughness**[nm Sq] | | 0.6 | 0.564 |
| | **Right surface micro-roughness** [nm Sq] | | 0.6 | 0.510 |
| | **Concave surface micro-roughness** [nm Sq] | | 0.6 | 0.472 |
| | **Concentricity of concave surface and diameter** [mm] | | 0.5 | 0.4 |

## Measuring methods

Measurement was the most important part of the production process. All measurements of mechanical dimensions were performed on a Mitutoyo Legex 744 3D coordinate measuring machine (CMM).

The first task was to suppress various side effects when measuring angles. Because it was essential to achieve very high accuracy in the production of the spacer, it was crucial to design a clamping system and establisha measurement process. In the first step, it was necessary to measure the angular values during grinding and angulation of surfaces. The Mitutoyo has a very precise table. This table was used as a zeroing and leveling surface for the measured space. The prism was placed and fixed on this table. Specifically, the angular relationships between the front surfaces and the side surfaces were measured.

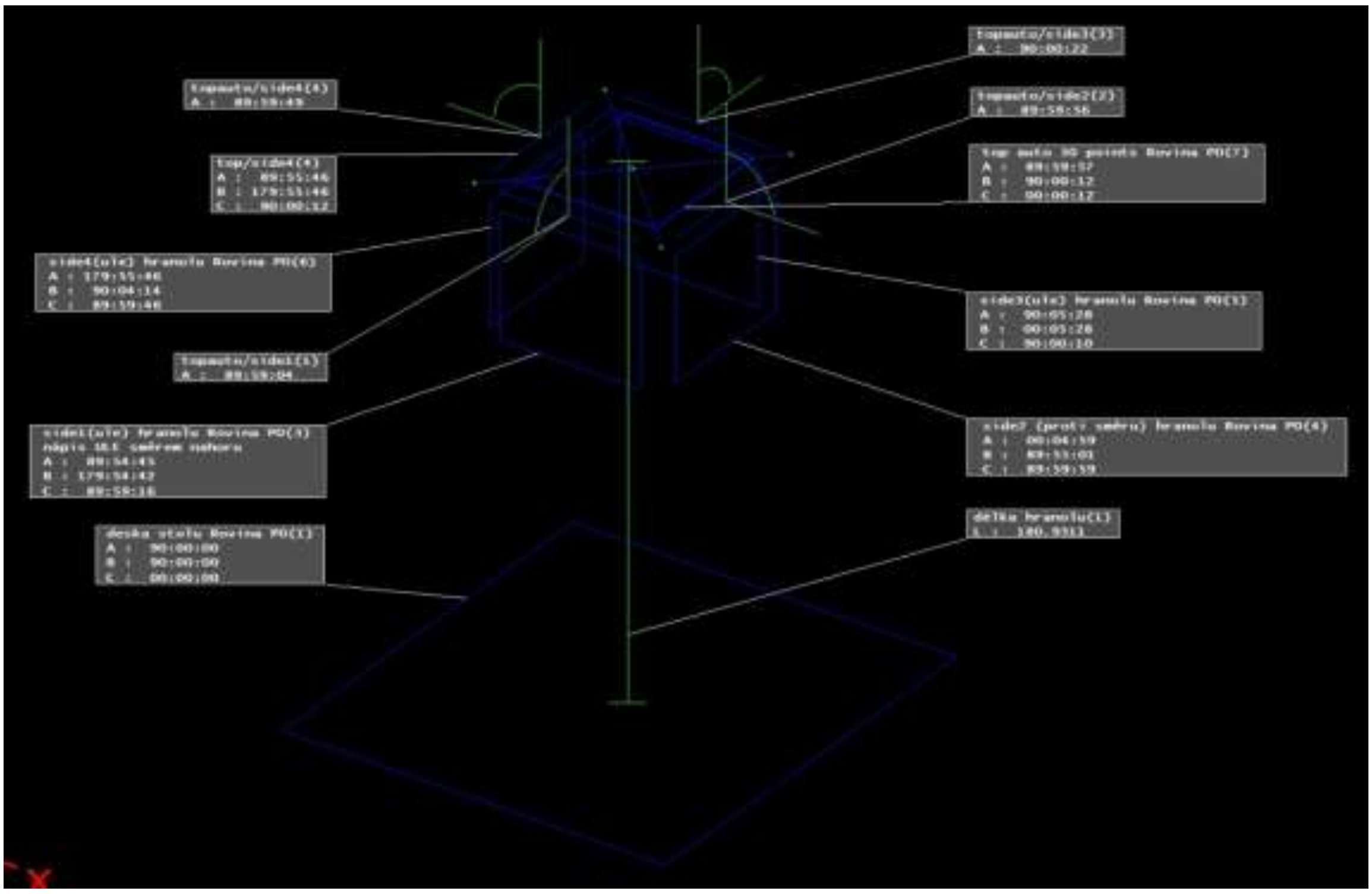

Figure 4. Illustration of a spacer model when measuring surface angles

Another interesting issue was the measurement of the drilled cavity. This cavity has a length of 181 mm. In order to be able to measure this cavity, it was necessary to use a probe whose measuring ball has a larger diameter than the diameter of the probe shaft. The Legex 744 uses a ruby ball probe for measuring. Mitutoyo has a large number of probes of various shapes and lengths in its portfolio. In our case, a probe with a length of 90 mm and a ball diameter of 8 mm was used.The measurement procedure itself then consisted of two mainsteps. First, the entire prism had to be focused to create a model. Subsequently, the cavity measurement was set in the automatic mode. The measurements were performed in circles with a distance of 10 mm, first on one side and then on the other in the range of 0–80 mm.

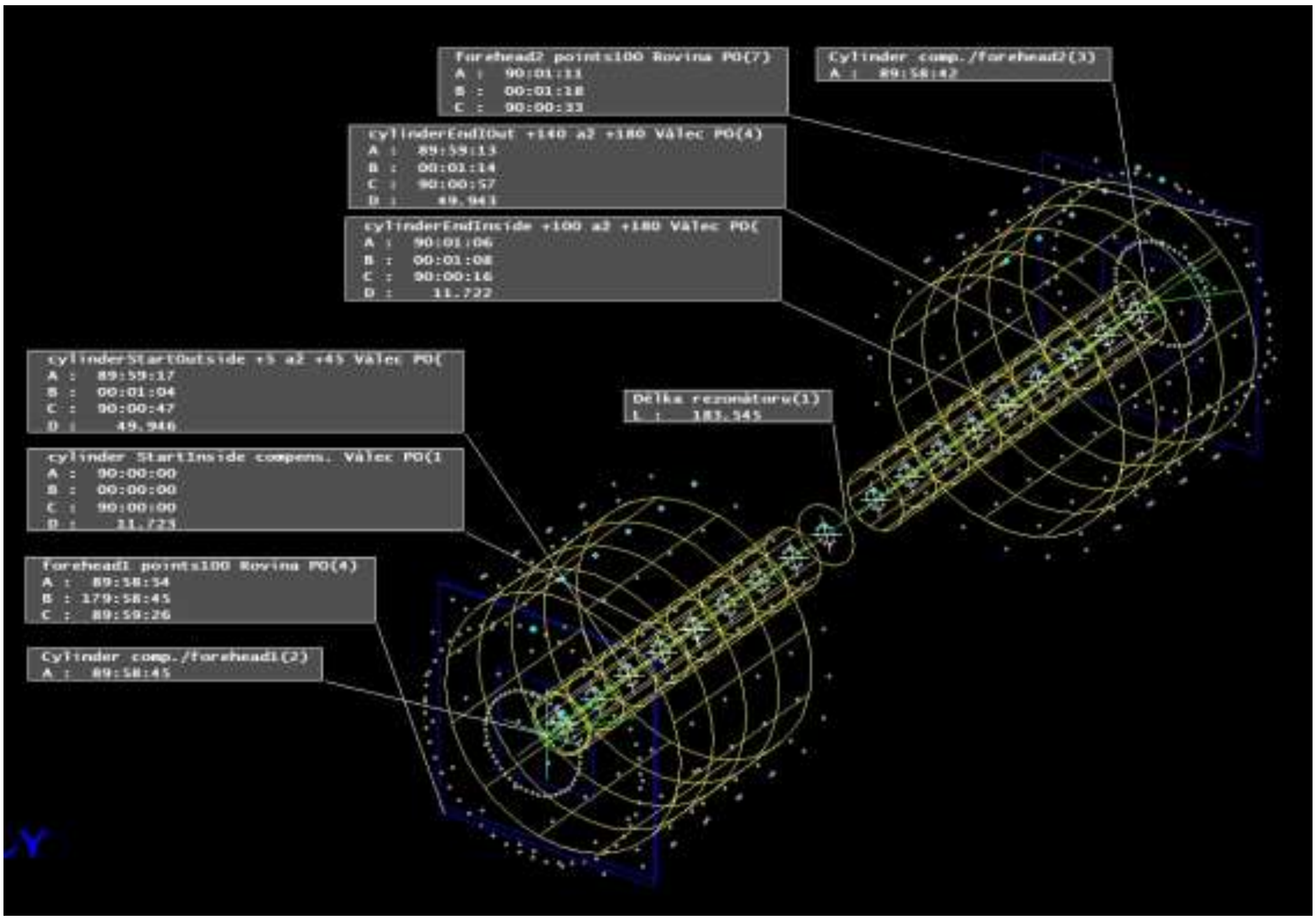

Figure 5. Illustration of a spacer model when measuring a cavity and a circumferential cylinder

Roundness was then calculated for each circle. Then the circles were arranged in a cylinder and the cylindricality of the cavity was evaluated, as was the flatness of the cavity axis. Because all the other spacer parameters apply to the cavity, a thorough examination of its geometric parameters was required. After measuring the cavity, the spacer was rounded and etched. The cavity was measured again. The concentricity of the cavity with the circumference and the perpendicularity of the end faces to the cavity axis were checked.

After all the parts were produced, the spacer was placed on a prism. Attachments were glued to the edges of the prism. The washers weremade exactly for the specific diameter of the mirrors and defined their central coaxiality with the spacer. In the next step, the mirrors and the compensating rings were placed on the spacers and pressed against the spacer (the outside of the mirror). Thanks to the precision of the planar surfaces, the mirror attaches to the front surface of the spacer. The resonator was slowly heated to 200 °C. Thus, the strength of all joints was verified.

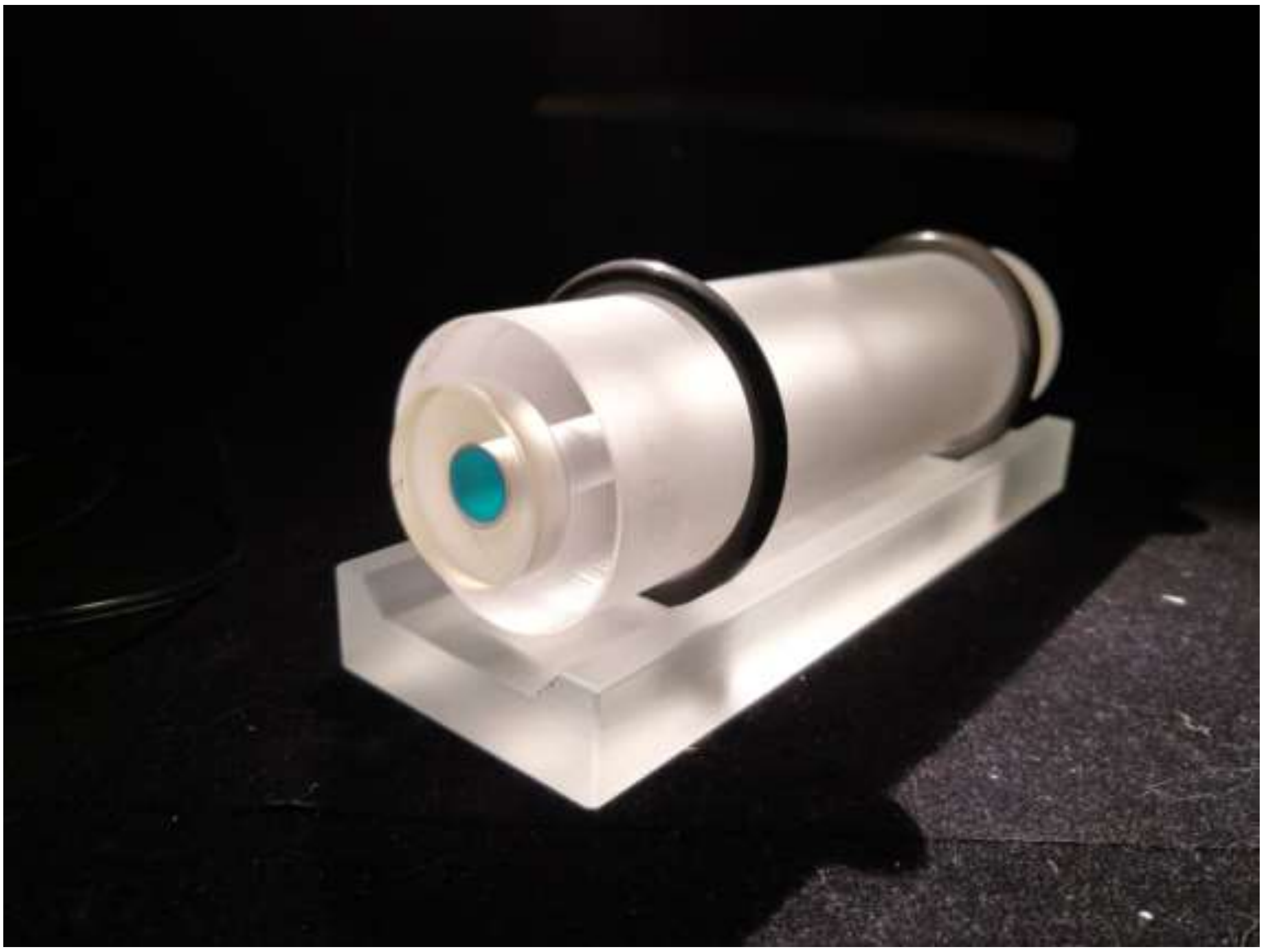

Figure 6. Photo of the resonator before fixing the stabilizing rings

## Conclusion

A manufacturing process of a Fabry–Perot type laser resonator was developed. The procedure was tested on a test piece and, subsequently, a production piece was fabricatedby this method. The resonator was made for a wavelength of 1540nm. The length of the spacer (distance of the mirrors) is 180 mm. All the mechanical elements of the resonator are made of ULE material, while optical elements are made of fused silica. During production, a very high angular accuracy and parallelism of the spacer faces, 5 arcsec, was achieved. A procedure for drilling long holes was developed so that a hole with a diameter of 12 mm and a length of 180 mm could be drilled into the manufactured resonator. On a length of 180 mm, a coaxiality of the inlet holes of 0.02 mm and a cylindricality of 0.014 mm were achieved. The whole assembly is designed and manufactured for work in a vacuum environment with a temperature of 31 °C and the space will be heated to this temperature. In the future, the goal will be the production of resonators featuring a spacer length of up to 500 mm.

## Acknowledgement

This work has been carried out within the framework of the Partnership for Excellence in Superprecise Optics project (Reg. No. CZ.02.1.01/0.0/0.0/16_026/0008390) and co-funded from European Structural and Investment Funds.

## References

[1] Zhao, Y. N., Zhang, J., Stuhler, J., Schuricht, G., Lison, F., Lu, Z. H., &Wang, L. J. (2010). Sub-Hertz frequency stabilization of a commercial diode laser. *Optics Communications*, *283*(23), 4696-4700.

[2] Ludlow, A. D., Boyd, M. M., Ye, J., Peik, E., & Schmidt, P. O. (2015). Optical atomic clocks. *Reviews of Modern Physics*, *87*(2), 637.

[3] Predehl, K., Grosche, G., Raupach, S. M. F., Droste, S., Terra, O., Alnis, J., & Schnatz, H. (2012). A 920-kilometer optical fiber link for frequency metrology at the 19$^{th}$decimal place. *Science*, *336*(6080), 441-444.

[4] ARNOLD, William R. Study of small variations of coefficient of thermal expansion in Corning ULE glass. In: *Optical Materials and Structures Technologies*. International Society for Optics and Photonics, 2003. p. 28-37.

[5] Alnis, J., Matveev, A., Kolachevsky, N., Udem, T., &Hänsch, T. W. (2008). Sub–hertz line width diode lasers by stabilization to vibrationally and thermally compensated ultra low-expansion glass Fabry-Pérotcavities. *PhysicalReview A*, *77*(5), 053809.

[6] Legero, T., Kessler, T., &Sterr, U. (2010). Tuning the thermal expansion properties of optical reference cavities with fused silica mirrors. *JOSA B*, *27*(5), 914-919.

[7] Fox, R. W. (2009). Temperature analysis of low–expansion Fabry–Perot cavities. *Optics express*, *17*(17), 15023-15031.

[8] Xu, G., Zhang, L., Chen, L., Jiao, D., Liu, J., Jiang, C., ... &Liu, T. (2018, April). Study on the effect of ULE rings on the vibration sensitivity of horizontal ultra-stable optical cavities. In *2018 European Frequency and Time Forum (EFTF)* (pp. 389-391). IEEE.